\documentclass[a4paper,11pt]{article}
\pdfoutput=1
\usepackage{jheppub}
\usepackage[T1]{fontenc}
\usepackage{graphicx,graphics}

\def\2tvec#1#2{ \left( \begin{array}{c}
#1  \\
#2  \\
\end{array} \right)}%
\def\mat2#1#2#3#4{ \left( \begin{array}{cc}
#1 & #2 \\
#3 & #4 \\
\end{array} \right) }%
\def\Mat3#1#2#3#4#5#6#7#8#9{ \left( \begin{array}{ccc}tri-bimaximal
#1 & #2 & #3 \\
#4 & #5 & #6 \\
#7 & #8 & #9 \\
\end{array} \right) } %
\def\Mat3#1#2#3#4#5#6#7#8#9{ \left(
\begin{array}{ccc}
#1 & #2 & #3 \\
#4 & #5 & #6 \\
#7 & #8 & #9 \\
\end{array} \right) }

\def\3tvec#1#2#3{ \left( \begin{array}{c}
#1  \\
#2  \\
#3  \\
\end{array} \right)}

\def\4tvec#1#2#3#4{ \left( \begin{array}{c}
#1  \\
#2  \\
#3  \\
#4  \\
\end{array} \right)}

\def\hbar{\hspace{1mm}\bar{}\hspace{-1mm}h}

   \def\nn{\nonumber}

\newcommand{\balg}{\begin{align}}
\def\bea{\begin{eqnarray}}
\def\eea{\end{eqnarray}} \newcommand{\be}{\begin{eqnarray}}
\newcommand{\ee}{\end{eqnarray}}

\title{Modular $A_4$ Symmetry With Three-Moduli and Flavor Problem}

\author{
{\textbf{Mohammed Abbas$^{1,2}$, Shaaban Khalil$^3$}}}

\affiliation{$^1$ Physics Department, Faculty of Science and Arts-Tabarjal, Jouf University, Al-Jouf, KSA.}
\affiliation{$^2$ Physics Department, Faculty of Sciences, Ain Shams University,
Abbassiyah 11566, Cairo, Egypt.}

\affiliation{$^3$Center for Fundamental Physics, Zewail City of Science and
Technology, 6th of October City, Giza 12578, Egypt.}

\emailAdd{maabbas@ju.edu.sa}
\emailAdd{skhalil@zewailcity.edu.eg}

\abstract{ 
The modular $A_4$ symmetry with three moduli is investigated. We assign different moduli to charged leptons, neutrinos, and quarks. We analyze these moduli at their fixed points where a residual symmetry exists. We consider two possibilities for right-handed neutrinos. First, they are assumed to be singlets under  the modular symmetry. In this case, we show that the lepton masses and mixing can be obtained consistently with experimental observations. Second, they are assigned non-trivially under modular symmetry. We emphasize that a small deviation from their fixed point is required in this case. Finally, the quark masses and mixing are generated correctly around the fixed point of their modulus. In our analysis, we only consider the simple case of weight 2.}

\begin{document}

\maketitle \flushbottom

\section{Introduction}

The origin of the quark and lepton masses, as well as their mixing, is one of the most important unsolved problems in particle physics. The experimental results of the fermion masses and mixings cannot be interpreted by the Standard Model (SM). Thus, they are providing the most reliable cue for new physics beyond the SM. The charged fermion masses and mixing angles in the SM are derived from arbitrary Yukawa couplings. Therefore, it is impossible to explain the enormous mass ratios between generations:
$m_u \ll m_c \ll m_t$; $m_d \ll m_s \ll m_b$; $m_e \ll m_\mu \ll m_\tau$ , as well as the large mass splitting within the third generation: $ m_\tau \sim m_b \ll m_t$.

Furthermore, the measured neutrino masses demonstrated the requirement for a method other than the conventional Yukawa couplings to explain their tiny values. Type I \cite{typeI} and Inverse seesaw \cite{inverseseesaw} are two intriguing examples for this mechanism. In addition, the quark mixing matrix is almost diagonal while the mixing of leptons (neutrinos) are quite large except the 13 element. The SM is unable to address all of this existing experimental data, and  discrete symmetries are sought to be taken into consideration.  However, they frequently involve large number of arbitrary parameters and complicated scalar sectors, which lessens their predictability.

Recently, finite modular groups $\Gamma_N$, inspired by string theory and extra dimensions compactified on orbifolds, have been proposed to explain the aforementioned flavor features \cite{modulargroups, Feruglio:2017spp}. In this framework, no flavons are required to be introduced, and the only source of modular (flavour) symmetry breaking is the Vacumme Expectation value (VEV) of the moduli fields. String theory has a rich moduli structure in general, and three moduli are naturally obtained in orbifold compactification \cite{Bailin:1999nk}.  It is worth noting that in string theory, moduli refer to the set of scalar fields that describe the shape and size of extra-dimensions. For the sake of simplicity, one modulus with equal contributions from all moduli may be assumed.
This was employed in the analysis of fermion masses and mixing from modular symmetries.
It turns out that with only one modulus $\tau$, it is not possible to account for fermion masses and mixing unless severe fine-tuning assumptions are made. For example, in the case of modular group $\Gamma_2 \sim S_3$, five observables must be determined in terms of only two parameters: $\tau$ and the ratio of two free parameters in the Lagrangian, which is a very difficult task.

The modular groups $\Gamma_N$ are isomorphic to some finite permutation groups. Most of the literatures assumed modular invariant models with only one modulus. For instance, $\Gamma_2\cong S_3$ \cite{Kobayashi:2018vbk, Okada:2019xqk,Du:2020ylx, Xing:2019edp}, $\Gamma_3\cong A_4$ \cite{Kobayashi:2018scp, Okada:2018yrn, Ding:2019zxk, Gui-JunDing:2019wap, Nomura:2019yft, Nomura:2019xsb, Asaka:2019vev, Abbas:2020qzc, Okada:2021qdf, Gunji:2022xig, Du:2022lij, Okada:2020ukr, Feruglio:2021dte, Novichkov:2021evw, Kikuchi:2022svo, Hoshiya:2022qvr}, $\Gamma_4\cong S_4$ \cite{Penedo:2018nmg, Kobayashi:2019mna,Qu:2021jdy,King:2021fhl,Liu:2020akv,Okada:2019lzv, Kobayashi:2019xvz, Wang:2019ovr} and $\Gamma_5\cong A_5$ \cite{Novichkov:2018nkm, Ding:2019xna, Criado:2019tzk}.
Some attempts to consider two moduli with different VEVs of fixed point values for charged lepton and neutrino sectors have been studied using the concept of modular residual symmetries \cite{Novichkov:2018yse, Novichkov:2018ovf}. Other attempt used multiple modular groups that have been broken effectively into a single modular group with multi moduli has been discussed in \cite{deMedeirosVarzielas:2019cyj}.

In this paper, we adopt the scenario of three moduli with modular group $\Gamma_3 \sim A_4$. This would allow us to assign different moduli to quark and lepton sectors.
More specifically, we conduct our analysis at the fixed points of these moduli. At these points the modular group $A_4$ is broken down to subgroups $Z_2$ or $Z_3$. Although, these subgroups are necessary for obtaining mixing matrices that are close to the observed ones, they should be approximated in order to get consistent results. We may choose to break the
residual symmetries by very small deviation from the moduli fixed points. It is worth mentioning that, thus far, fixing the SM fermions in extra-dimensions at specific fixed points with given Yukawa couplings in terms of moduli fields is just an assumption. All extra-dimensional theories are still in their early stages of development. However, it is critical to investigate the concepts proposed by these theories and examine their phenomenological implications.

This paper is organized as follows. Section II provides a brief overview of modular symmetry, with an emphasis on the three moduli scenario. Section III is devoted to introducing $A_4$ modular invariance. In section IV, we study the modular forms near the fixed points and how small is the deviation from the fixed points. In section V, detailed results of charged lepton and neutrino masses and mixing are calculated. In Section VI we investigate quark masses and mixing. Finally, our conclusions are given in Section VII.

\section{Modular Symmetry with Three Moduli}
Toroidal compactification is one of the simple  compactifications of extra dimensions. The two-dimensional tours $T^2$ is constructed by $R^2/\Lambda$, where $\Lambda$ is two-dimension lattice spanned by the two lattice vectors: $\alpha_1=2\pi R$ and $\alpha_2 = 2 \pi R \tau$ \cite{Kobayashi:2017dyu}, with $R$ is a real parameter and $\tau$ is a complex modulus parameter. The same lattice can be spanned by another basis given by
\begin{equation}
\left(
\begin{array}{c}
 \alpha'_1\\
\alpha'_2  \end{array} \right) = \left(
\begin{array}{cc}
a & b \\
c & d  \end{array} \right)  \left(
\begin{array}{c}
 \alpha_1\\
\alpha_2  \end{array} \right) ,
\end{equation}
where $a,b,c,d$ are integers that satisfy the relation $ad - bc=1$. This is known as $SL(2,Z)$ transformation. The modulus $\tau = \alpha_2/\alpha_1$ transforms as
\be
\tau \to \gamma(\tau) = \frac{a \tau + b}{c \tau + d}.
\ee
The group $SL(2,Z)$ is called modular group. It is generated by two elements $S$ and $T$, which are given by
\be
S=  \left( \begin{array}{cc}
0 & 1 \\
-1 & 0  \end{array} \right), ~~~ ~~~ ~~  T= \left(\begin{array}{cc}
1 & 1 \\
0 & 1  \end{array} \right).
\ee
The matrices $S$ and $T$ satisfy the relations $S^2= (ST)^3= I$, and their lead to the transformation
\be
S: \tau \to - \frac{1}{\tau}, ~~~ ~~~ ~~  T: \tau \to \tau +1.
\ee

The modular group $SL(2,Z)$ is generalized to group $\Gamma(N), N=1,2,3,...$, defined as follow:
\be
\Gamma(N) = \Big\lbrace    \left(
\begin{array}{cc}
a & b \\
c & d  \end{array} \right) \in SL(2,Z), ~  \left(
\begin{array}{cc}
a & b \\
c & d  \end{array} \right)  =  \left(
\begin{array}{cc}
1 & 0 \\
0 & 1  \end{array} \right) ({\rm mod}~N) \Big\rbrace,
\ee
where $\Gamma = \Gamma(1)=SL(2,Z)$. The group $\Gamma(N)$ acts on $\tau$ , varying in the upper-half complex plane ${\cal H} = {\rm Im}(\tau) >0$.
For $N=1,2$, the projective principal congruence subgroups are defined as $\bar{\Gamma}(N) = \Gamma(N)/\{I,-I\}$, and for $N>2$ as $\bar{\Gamma} (N) = \Gamma(N)$ because the element $-I$ does not belong to $\Gamma(N)$ for $N>2$.

The quotient groups $\Gamma_N =\bar{\Gamma}(1)/\bar{\Gamma}(N)$ are usually called finite modular
groups, which can be  obtained from $\bar{\Gamma}(1)$ by imposing the condition: $T^N = 1$ \cite{Feruglio:2017spp}.
The groups $\Gamma_N$ with $N = 2, 3, 4, 5$ are isomorphic to the permutation groups $S_3, A_4, S_4$ and
$A_5$ respectively \cite{deAdelhartToorop:2011re}. It is worth noting that in string theory, the compactification of six extra dimensions on a toroidal orbifold $T^6/\Gamma$, where $\Gamma$ is a finite isometry group, such as $\Gamma=Z_N$, which acts with fixed point, implies the existence of three moduli $\tau_i$, $i=1,2,3$ \cite{Baur:2019kwi}, rather than just one modulus, as is commonly assumed in the literature. The existence of discrete modular symmetry on the toroidal background such as $T^6/(Z_2×Z_2)$ was generalized to Calabi-Yau space including multiple moduli fields in \cite{Ishiguro:2021ccl}. In this case, the modular transformation is defined by
\be
\tau'_i = \frac{a_i \tau_i + b_i}{c_i \tau_i + d_i},
\ee
with
\be
S: \tau_i \to - \frac{1}{\tau_i}, ~~~ ~~~ ~~  T: \tau_i \to \tau_i +1.
\ee
The chiral superfield $\phi^\alpha$, as a function of $\tau_i$ transforms as
\be
\Phi^\alpha(\tau_i)  \to (c_i \tau_i +d_i)^{-k_\alpha} \rho(\gamma) \Phi^\alpha,
\ee
where $k_\alpha \in Z$ is the modular weight and $\rho$ is arbitrary representation of $\Gamma$.  For an N = 1 supersymmetric model, the superpotential is given by
\be
W = \frac{1}{6} Y_{\alpha\beta\gamma}(\tau_i) \Phi^\alpha  \Phi^\beta  \Phi^\gamma.
\ee
Because $W$ is required to be invariant under all modular transformation, the Yukawa couplings should be modular forms of weight
\bea k_Y = k_\alpha + k_\beta + k_\gamma,\label{modularinvariancecondition}
\eea
 {\it i.e.},
\be
Y_{\alpha\beta\gamma}(\gamma \tau_i)=  (c_i \tau_i +d_i)^{k_Y} \rho (\gamma)  Y_{\alpha\beta\gamma}(\tau_i).
\ee

In our analysis, we use the freedom of having three moduli with different VEVs and try to assign one modulus to each sector. In particular, $\tau_\ell$, $\tau_\nu$, and $\tau_q$, for charged lepton, neutrino, and quark sectors, respectively.

It is important to emphasize that in our study, we treat the charged leptons, neutrinos,
as well as up- and down-type quarks as unified left-handed doublets (\textbf{L} and \textbf{Q}, respectively).
This approach stems from the observation that the compactification scale, where fixed
points in the extra dimensions are identified, is much higher than the electroweak symmetry
breaking scale. Another noteworthy aspect to consider is that in this scenario, the Yukawa
couplings are generated through the overlap of the profiles of the relevant fields in their
interaction at specific fixed points, each corresponding to a certain value of a moduli VEV.
Consequently, we can have a Yukawa term denoted as \textbf{$Y(\tau_l)$} for the interaction involving the
fields 
\textbf{$E$},\textbf{$H_d$}, and \textbf{$L$}, while a separate interaction involving the fields \textbf{$N$},\textbf{$H_u$}, and \textbf{$L$} occurs at a distinct fixed point represented by $\tau_{\nu}$, resulting in the corresponding Yukawa term \textbf{$Y(\tau_{\nu})$}.

 In fact, finding a mechanism that stabilizes the moduli fields at specific VEVs, at or near fixed points, in order to obtain a consistent physical theory, is an open question in string theory. The details of moduli stabilisation are beyond the scope of this paper. For more details about the moduli stabilization near the fixed points see \cite{Kobayashi:2020uaj, Abe:2020vmv}.
As stated in the introduction, accounting for fermion masses and mixing simultaneously requires the free parameters in the Lagrangian to be fine-tuned. However, it was difficult to account for both lepton and quark mixing at the same time, so an attempt was made to focus solely on the lepton sector.

\section{$A_4$ modular invariance and residual symmetries}
The group $A_4$ has one triplet representation $\textbf{3}$ and three singlets $\textbf{1}, ~\textbf{1}^{\prime}$ and $\textbf{1}^{\prime\prime}$ and is generated by two elements $S$ and $T$ satisfying the conditions
\bea
S^2=T^3=(ST)^3=\textbf{1}.
\eea
The modular form of level 3 has the form
$$f_i(\gamma(\tau))=(c\tau+d)^{2k} \rho_{ij}(\gamma) f_j(\tau), ~~\gamma \in \Gamma(3).$$ The modular form of weight 2 and level 3 transforms as a triplet and is given by $Y_3^{(2)}=(y_1, y_2, y_3)$, \cite{Feruglio:2017spp} where :
\bea
y_1(\tau) &=& \frac{i}{2\pi}\left[ \frac{\eta'(\tau/3)}{\eta(\tau/3)}  +\frac{\eta'((\tau +1)/3)}{\eta((\tau+1)/3)}
+\frac{\eta'((\tau +2)/3)}{\eta((\tau+2)/3)} - \frac{27\eta'(3\tau)}{\eta(3\tau)}  \right], \nonumber \\
y_2(\tau) &=& \frac{-i}{\pi}\left[ \frac{\eta'(\tau/3)}{\eta(\tau/3)}  +\omega^2\frac{\eta'((\tau +1)/3)}{\eta((\tau+1)/3)}
+\omega \frac{\eta'((\tau +2)/3)}{\eta((\tau+2)/3)}  \right] ,\nonumber \\
y_3(\tau) &=& \frac{-i}{\pi}\left[ \frac{\eta'(\tau/3)}{\eta(\tau/3)}  +\omega\frac{\eta'((\tau +1)/3)}{\eta((\tau+1)/3)}
+\omega^2 \frac{\eta'((\tau +2)/3)}{\eta((\tau+2)/3)} \right]\,.\label{Y_forms}
\eea
where $\omega=e^{2i\pi/3}$ and the Dedekind eta-function $\eta(z)$ is defined as
\bea
\eta(\tau)=q^{1/24} \prod_{n =1}^\infty (1-q^n), \qquad  q=e^{2\pi i\tau}.\,
\eea
The $q$-expansion of $y_i(\tau)$ are:
\bea
y_1(\tau)&=&1+12q+36q^2+12q^3+...\nn\\
y_2(\tau)&=&-6q^{1/3}(1+7q+8q^2+...)\nn\\
y_3(\tau)&=&-18q^{2/3}(1+2q+5q^2+...)~~~.\label{qexpansion}
\eea
We will use the basis where the generators of $A_4$ in triplet representation are
\bea
S=\frac{1}{3}\left(
               \begin{array}{ccc}
                 -1 & 2 & 2 \\
                 2 & -1 & 2 \\
                 2 & 2 & -1 \\
               \end{array}
             \right), ~~~~T=\left(
                              \begin{array}{ccc}
                                1 & 0 & 0 \\
                                0 & \omega & 0 \\
                                0 & 0 & \omega^2 \\
                              \end{array}
                            \right).
                            \eea
There are independent fixed points,
\bea
\tau_1  &=& e^{\frac{i 2 \pi}{3}}=\frac{-1}{2}+i\frac{\sqrt{3}}{2},\nonumber\\
\tau_2 &=& i,\nonumber\\
\tau_3 &=& i\infty.
\eea
Because these fixed points are invariant under modular transformations, any other fixed point is equivalent to the above fixed points. For example, the fixed point $\tau_{1^{\prime}}=\frac{1}{2}+i\frac{\sqrt{3}}{2}$ is a result of the action of $T$ on $\tau_1$, $\tau_{2^{\prime}}=-0.5+ 0.5 i$ is equivalent to $\tau_2$ when transformed by $ST$. The $TST$ transformation of $\tau_2$ can also be used to obtain the point $\tau_{2^{\prime\prime}}=0.5 + 0.5 i$. Any of the moduli $\tau_\ell$, $\tau_\nu$, and $\tau_q$, could have one of the above fixed point values.

It is worth noting that the modular symmetry is broken down to a residual symmetry at the fixed points, which may be important if specific fermion mixing is required. Specifically, the point $\tau_1$ is invariant under $ST$ transformation, thus at $\langle \tau \rangle=\tau_1$, the modular group $A_4$ is broken to its subgroup $Z_3=\{I, ST, (ST)^2\}$. In this case, the fermion mass matrices are invariant under $ST$ transformation. Whereas $\tau_2$ is invariant under $S$ transformation, so the modular group $A_4$ is broken to its subgroup $Z_2=\{I, S \}$ and the fermion mass matrices are invariant under $S$ transformation. The point $\tau_{2^{\prime}}=-0.5+0.5i$ is invariant under $ST^2ST$ transformation and $A_4$ is broken to its subgroup $Z_2=\{I, ST^2ST\}$. The infinite point $\tau_3$ is invariant under $T$ transformation, thus at $\langle \tau \rangle=\tau_3$, the modular group $A_4$ is broken to its subgroup $Z_3=\{I, T, T^2\}$. It was proven in \cite{Ishiguro:2020tmo} that a fixed point with the residual $Z_3$ symmetry
in the SL(2; Z) fundamental region is statistically favored in the string landscape from the viewpoint of modular flavor symmetries.

We consider that the modular flavor symmetry is broken differently in charged lepton, neutrino and quark sectors into different residual symmetries via the assignment with different moduli.

\section{Modular forms near fixed points}
In order to determine whether a value of a modulus is close to fixed point or not, we perform a Taylor expansion of the modular forms around the fixed points by small parameter $\epsilon$ and find the value of $\epsilon$ that allowing us to safely ignore higher orders.
\subsection{Modular forms close to $\tau=i$}
We parameterize $\tau$ as
\bea
\tau= i + \epsilon,
\eea
where $\epsilon\ll 1$. If $\tau$ is close to $i$, we can perform linear fit of $\epsilon$ as
\bea
\frac{y_2(\tau)}{y_1(\tau)}=(1+\epsilon_1)(1-\sqrt{3}), ~~~~ \frac{y_3(\tau)}{y_1(\tau)}=(1+\epsilon_2)(-2+\sqrt{3}), \label{ratios}
\eea
where
\bea
\epsilon_1=\frac{\epsilon_2}{2}=2.06~ i\epsilon.
\eea
For $\mid \epsilon \mid \leq 0.06$, the approximations of the modular form ratios in Eq. \ref{ratios} are in agreement with the exact ratios within $1 \% $.

\subsection{Modular forms close to $\tau=\omega$}
We parameterize $\tau$ as
\bea
\tau= \omega + \epsilon,
\eea
where $\epsilon\ll 1$. If $\tau$ is close to $\omega$, we can perform linear fit of $\epsilon$ as
\bea
\frac{y_2(\tau)}{y_1(\tau)}=\omega (1+\epsilon_1), ~~~~ \frac{y_3(\tau)}{y_1(\tau)}=-\frac{\omega^2}{2}(1+\epsilon_2), \label{ratios2}
\eea
where
\bea
\epsilon_1=\frac{\epsilon_2}{2}=2.17~ i\epsilon.
\eea
For $\mid \epsilon \mid \leq 0.06$, the approximations of the modular form ratios in Eq. \ref{ratios2} are in agreement with the exact ratios within $0.05 \% $.

\section{Lepton masses and mixing}

In this section, we apply the above-mentioned formalism to SM extensions involving right-handed neutrinos and the type I seesaw mechanism. In contrast to most flavour symmetric models, our model does not include flavons or extra discrete symmetries.
According to the modular invariance condition Eq.(\ref{modularinvariancecondition}), we fix the modular weights so that the following relations are satisfied:
\bea
k_L+k_{H_d}+k_E&=&2,\nonumber\\
k_L+k_{H_u}+k_N&=&2.
\eea

The assignments under $A_4$ and modular weights for the lepton and Higgs fields are shown in Table (\ref{assignment 1}).
As we mention above, we assume that the model contains multiple moduli, each for a different sector. Due to the symmetric nature of the neutrino mass matrix, the residual symmetry in neutrino sector can be $Z_2$ symmetry, while it can be $Z_3$ in charged lepton sector \cite{Lam:2011ag}. Thus we set $\tau_{\ell}=\tau_1$ or $\tau_3$, while $\tau_{\nu}$ can acquire any of the values $\tau_2=i$, $\tau_{2^{\prime}}=-0.5+ 0.5 i$ or $\tau_{2^{\prime\prime}}=0.5+ 0.5 i$.

\begin{table}[h!]
  \centering
  \begin{tabular}{|c|c|c|c|c|c|c|c|c|c|}
  \hline
  fields & L & $E^c_1$ &$E^c_2$ &$E^c_3$&$N^c$&$ H_d$ &$ H_u$  \\
   \hline
  $A_4$ & 3 & 1 & $1^{\prime\prime}$ &$1^{\prime}$&3  & 1 &1 \\
   \hline
  $k_I$ & $k_L$ & $k_{E_1}$ & $k_{E_2}$ & $k_{E_3}$& $k_N$  & 0&0  \\
  \hline
\end{tabular}
  \caption{Assignment of flavors under $A_4$ and the modular weight $k_I$}\label{assignment 1}
\end{table}

\subsection{Charged Lepton sector}

We start with charged lepton sector. The modular $A_4$ invariant superpotential of this sector is given by
\bea
W_\ell&=&\lambda_1 E_1^c H_d\Big(L \otimes Y_3^{(2)}(\tau_{\ell})\Big )_1 +\lambda_2 E_2^c H_d \Big(L \otimes Y_3^{(2)}(\tau_{\ell})\Big)_{1^{\prime}}+\lambda_3 E_3^c H_d\Big(L \otimes Y_3^{(2)}(\tau_{\ell})\Big)_{1^{\prime\prime}}.
\label{chargedleptonsuperpot.}
\eea
%
From this equation, one can write the charged lepton mass matrix as
\bea
m_{\ell}&=&v_d \left(
                      \begin{array}{ccc}
                        \lambda_1 & 0 & 0 \\
                        0 & \lambda_2 & 0 \\
                        0 & 0 & \lambda_3 \\
                      \end{array}
                    \right)\times \left(
      \begin{array}{ccc}
        y_1 & y_3 & y_2 \\
        y_2 & y_1 & y_3 \\
        y_3 & y_2 & y_1 \\
      \end{array}\right).
      \eea
It is convenient to use the Hermitian matrix $M_{\ell}=m_{\ell}^{\dagger} ~m_{\ell}$ which can be diagonalized by a single unitary matrix:
$$M_{\ell}^{diag}=U_{\ell}^{\dagger}M_{\ell} U_{\ell} .$$
If $\tau_{\ell}=\tau_1,$ then the charged lepton Hermitian mass matrix, $M_{\ell}$, is invariant under $ST $ transformation, $$(ST)^{\dagger}~M_{\ell}~ ST=M_{\ell}.$$
Both $ST$ and $M_{\ell}$ are diagonalized by the Unitary matrix:
\bea
U_{\ell}=\frac{1}{3}\left(
      \begin{array}{ccc}
        2 \omega & -\omega & -2\omega \\
        2 \omega^2 & 2\omega^2 & \omega^2 \\
        -1 & 2 & -2 \\
      \end{array}
    \right).\label{chargedleptonmixing}
    \eea
We set the coupling ratios to be equal the experimental mass ratios, $\frac{\lambda_i}{\lambda_j}=\frac{m_{\alpha}}{m_{\beta}}$ where $i, j=1, 2, 3;~ \alpha, \beta= e, \mu, \tau$. The couplings may be permuted to obtain the experimental mixings, while the masses are invariant under the permutations. For instants, if $\frac{\lambda_2}{\lambda_1}=\frac{m_e}{m_{\tau}}=0.0003, ~~\frac{\lambda_3}{\lambda_1}=\frac{m_{\mu}}{m_{\tau}}=0.06$, the experimental charged lepton mass ratios are achieved.

If $\tau_\ell=\tau_3$, from Eq. \ref{qexpansion}, $y_1(\tau_3)=1,~ y_2(\tau_3)=0, y_3(\tau_3)=0$. consequently, the charged lepton mass matrix reads
\bea
m_{\ell}&=&v_d \left(
                      \begin{array}{ccc}
                        \lambda_1 & 0 & 0 \\
                        0 & \lambda_2 & 0 \\
                        0 & 0 & \lambda_3 \\
                      \end{array}
                    \right)
                    \eea
In this case the mixing is totally from neutrino sector.
\subsection{Neutrino sector}

Now we turn to neutrino sector.  As stated in the introduction, there could be more than one scenario for the assignments of right-handed neutrinos. We will start with singlet right-handed neutrinos under the modular symmetry ($ k_N =0$). Then we look at the case when the right-handed neutrinos are transformed non-trivially under modular symmetry. \\

\emph{\textbf{Model with $k_N=0$}}\\

Here we consider type-I seesaw mechanism, in which right-handed neutrinos are singlet under modular symmetry, hence we have
\bea
k_L=2,~ k_{E_1}=k_{E_2}=k_{E_3}=0,~ k_N =0.
\eea
The neutrino modular $A_4$ invariant superpotential is
\bea
W_{\nu}=g_1^{\prime}~
((N^c H_u L)_{3S}Y_3^{(2)}(\tau_{\nu}))_1+g_2^{\prime}~((N^c H_u L)_{3A}Y_3^{(2)}(\tau_{\nu}))_1+
 \Lambda~ N^c \otimes N^c,\label{nusuperpot.}
\eea
The neutrino mass matrices are
      \bea
 M_R=\Lambda\left(
                    \begin{array}{ccc}
                    1 & 0 & 0 \\
    0 & 0 & 1 \\
    0 & 1 & 0 \\
                    \end{array}
                  \right),~~ m_D &=& v_u\left(
  \begin{array}{ccc}
    2 g_1 y_1 & (-g_1+g_2) y_3 & (-g_1-g_2) y_2 \\
    (-g_1-g_2) y_3 & 2 g_1 y_2 & (-g_1+g_2) y_1 \\
    (-g_1+g_2) y_2 & (-g_1-g_2) y_1 & 2 g_1 y_3 \\
  \end{array}\right).\label{numassmatrix1}.
\eea
In the type-I seesaw, the neutrino mass matrix in the basis $(\nu_L, N^c)$ is given by
\bea
M=\left(
          \begin{array}{ccc}
            0 & m_D \\
            m_D^T &  M_R\\
            \end{array}
        \right).
        \eea
By diagonalization of this matrix, the light neutrino can be obtained as
\bea
m_{\nu}=-m_D {M_R}^{-1} m_D^T.
\eea
The overall parameter $v_u^2 g_1^2/\Lambda$, where $v_u = \langle H_u \rangle$, determines the scale of light neutrino masses and can be easily chosen to achieve the desired scale. For instant, we can set $v_u\sim {\cal O}~(10^2$ GeV), with  $g_1\sim{\cal O}(10^{-1})$ and $\Lambda\sim{\cal O}(10^{13}$) GeV or $g_1\sim{\cal O}(10^{-6}$) GeV and $\Lambda\sim{\cal O}(10^{5}$) TeV, to get the neutrino masses of order ${\cal O}(10^{-1}$) eV.

The neutrino mass matrix $m_{\nu}$ is complex and symmetric, so it is convenient to diagonalize the Hermitian matrix $M_{\nu}=m_{\nu}^{\dagger} m_{\nu}$,
\bea
M_{\nu}^{diag}=U_{\nu}^{\dagger}M_{\nu}U_{\nu} .
\eea
The lepton mixing $U_{PMNS}$ matrix is given by
\bea
U_{PMNS}=U_{\ell}^{\dagger} U_{\nu}.
\eea
The mixing angles can be calculated from the relations
\bea
Sin^2(\theta_{13})=|{(U_{PMNS})_{13}}|^2, ~Sin^2(\theta_{12})=\frac{|{(U_{PMNS})_{12}}|^2}{1-|{(U_{PMNS})_{13}}|^2}, ~Sin^2(\theta_{23})=\frac{|{(U_{PMNS})_{23}}|^2}{1-|{(U_{PMNS})_{13}}|^2}.
\eea
\begin{table}
\centering
\begin{tabular}{|c|c|c|c|c|c|c|c|c|c|}
  \hline
  & &$\frac{\Delta m^2_{12}}{(10^{-5}~\text{eV}^2)}$ & $\frac{|\Delta m^2_{23}|}{(10^{-3}~\text{eV}^2)}$ & $r=\frac{\Delta m^2_{12}}{|\Delta m^2_{23}|}$&$\theta_{12}/^{\circ}$ & $\theta_{23}/^{\circ}$ &$ \theta_{13}/^{\circ} $&$\delta_{CP}/\pi$ \\
   \hline
 Inverted Hierarchy& best fit& 7.39&-2.51&0.0294&33.82&49.8&8.65&1.57\\
 & $3\sigma$ range& 6.79-8.01 & -2.412-(-2.611) &0.026-0.033& 31.61-36.27 & 40.6-52.5& 8.27-9.03 &1.088-2  \\
  \hline
  Normal Hierarchy&best fit&7.39&2.525&0.0294&33.82&49.6&8.61&1.194\\
 & $3\sigma$ range& 6.79-8.01 & 2.427-2.625 &0.026-0.033& 31.61-36.27 & 40.3-52.4& 8.22-8.99 &0.694-2.177  \\
  \hline
\end{tabular}
\caption{3$\sigma$ range for neutrino mixings and mass difference
squares from \cite{Esteban:2018azc} for inverted hierarchy.}\label{recentdata}
\end{table}
The parameter $g_2/g_1$ is complex in general, so we can write it as
\bea
\frac{g_2}{g_1}=g e^{i \phi},
\eea
where $\phi$ is the relative phase of $g_1$ and $g_2$.
The best fit values and $3 \sigma$ ranges for the experimental results are summarized in Table (\ref{recentdata}), in which the neutrino mass squared differences are defined as
$$\Delta m_{12}^2=m_2^2-m_1^2, ~~~|\Delta m_{23}^2|=|m_3^2-(m_2^2+m_1^2)/2|.$$

If we consider $\tau_{\nu}=\tau_2=i$, the Hermitian Neutrino mass matrix $M_{\nu}$ is invariant under $S$ transformation, $$S^{\dagger}~ M_{\nu} ~S =M_{\nu}.$$ This leads to the scenario of strong hierarchy where one of the neutrino masses nearly vanishes while the other two masses are not degenerate contrary to the case of five dimensional non-renormalizamble operator studied in \cite{Novichkov:2018yse}.

We have four free parameters,$g$, $\phi$, $\frac{\lambda_1}{\lambda_3}$ and $\frac{\lambda_2}{\lambda_3}$  which can be tuned to obtain the observed lepton masses and mixings.

\begin{figure}[htb!]
\centering
\includegraphics[width=5cm,height=5cm,angle=0]{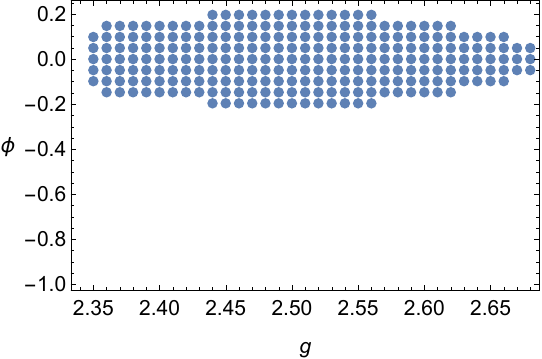}$~~~~~~$
\includegraphics[width=5cm,height=5cm,angle=0]{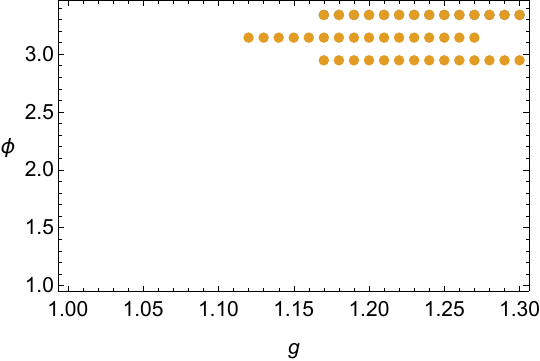}
\caption{The allowed values of $g$ and $\phi$ that yield to the normal hierarchy neutrino masses and observed mixing angles if (left) $\tau_{\ell}=e^{2\pi i/3},~\tau_{\nu}=i$ or $\tau_{\ell}=i\infty,~\tau_{\nu}=0.5+ 0.5i$, (right) $\tau_{\ell}=e^{2\pi i/3},~\tau_{\nu}=0.5+ 0.5i$ or $\tau_{\ell}=i \infty,~\tau_{\nu}=i$.}
\label{fig1}
\end{figure}

\begin{figure}[htb!]
\centering
\includegraphics[width=5cm,height=5cm,angle=0]{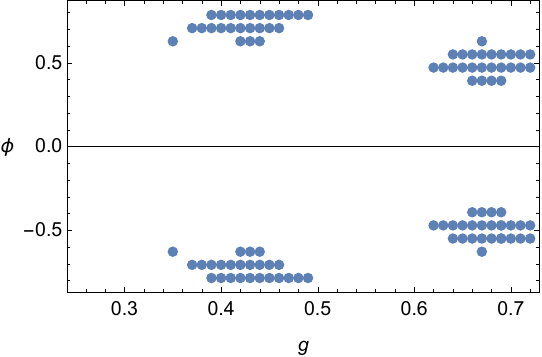}
\caption{The allowed values of $g$ and $\phi$ that yield to the inverted hierarchy neutrino masses and observed mixing angles if $\tau_{\ell}=e^{2\pi i/3},~\tau_{\nu}=i$ or $\tau_{\ell}=i \infty, ~\tau_{\nu}=0.5+ 0.5i$.}
\label{fig2}
\end{figure}

Note that because CP violation does not impose a severe constraint, and CP phases (Majorana and in mixing matrices) can be easily obtained in a modular symmetry scenario from the complex fixed points that represent the complex VEV of the flavon fields, this leads to a CP violation mechanism known as spontaneous CP violation. In this work, we chose to consider the real values of the PMNS mixing matrix entries while keeping the spontaneous CP for further analysis.

For the normal hierarchy case, as shown in Fig. \ref{fig1}, we can tune $g$ and $\phi$ together to get lepton masses and mixing within their experimental limits shown in Table \ref{recentdata}. For instance, we have the following benchmarks:
\begin{itemize}
  \item When $\tau_{\ell}=\tau_1,~\tau_{\nu}=\tau_2,$ or $\tau_{2^\prime},~\frac{\lambda_1}{\lambda_2}=0.0003, ~~\frac{\lambda_3}{\lambda_2}=0.06$, the observed data of mass ratios and mixing angles are achieved at $~g=[2.35, 2.68],~\phi=[ -0.07\pi,0.07\pi],~$. The same results can be obtained when $\tau_{\ell}=\tau_3,~\tau_{\nu}=\tau_{2^{\prime\prime}}$ with a permutation of $\lambda_i$ namely $~\frac{\lambda_2}{\lambda_3}=0.0003, ~~\frac{\lambda_1}{\lambda_3}=0.06$.
  \item If $\tau_{\ell}=\tau_1,~\tau_{\nu}=\tau_{2^{\prime\prime}},~ \frac{\lambda_1}{\lambda_2}=0.0003, ~~\frac{\lambda_3}{\lambda_2}=0.06$, the experimental values of the observed data occur at $g=[1.17, 1.3],~\phi=[0.93\pi,1.07\pi].$ One can get the same results when $\tau_{\ell}=\tau_3,~\tau_{\nu}=\tau_2$ or $\tau_{2^\prime}$ with the permutations of $\lambda_i$ as $~\frac{\lambda_2}{\lambda_3}=0.0003, ~~\frac{\lambda_1}{\lambda_3}=0.06$.
\end{itemize}

For the inverted hierarchy case, $m_2\gtrsim m_1\gg m_3$, if $\tau_{\ell}=\tau_1,~\tau_{\nu}=\tau_2$ or $\tau_{\nu}=\tau_{2^\prime}$, with $\frac{\lambda_1}{\lambda_3}=0.06, ~~\frac{\lambda_2}{\lambda_3}=0.0003$ or if $\tau_{\ell}=\tau_3,~\tau_{\nu}=\tau_{2^{\prime}}$ or $\tau_{\nu}=\tau_{2^{\prime\prime}}$~with $ \frac{\lambda_1}{\lambda_2}=0.06, ~~\frac{\lambda_3}{\lambda_2}=0.0003$, the allowed values of $g$ and $\phi$ that yield to observed fermion masses and mixing angles can be shown in Fig. \ref{fig2}. As shown from the figure, the average values of $\phi$ are $\pm \pi/4$.


The same results for normal and inverted hierarchies can be obtained via inverse seesaw mechanism when no modular forms coupled to both $M_R$ and $\mu_s$ matrices, where $M_R$ and $\mu_s$ are the mass matrices arisen by the mixing between $\nu_R$ and the singlet fermion $S$ and $S$ itself respectively.\\
In this model, the observed masses and mixings are arisen at exact fixed points $\tau_{\ell}=\tau_1,~\tau_3 ~\tau_{\nu}=\tau_2, ~\tau_{2^\prime},~\tau_{2^{\prime\prime}}$, so the observed lepton
masses and mixing are consequences of breaking the modular symmetry to $Z_3$ and $Z_2$ residual symmetries in charged leptons and neutrino respectively.\\

\emph{\textbf{Model with $k_N=1$}}\\

Now consider the type-I seesaw case, in which the right-handed neutrinos have non-vanishing weight. Here is an example of how leptons and neutrinos are weighted:
\bea
k_L=k_{E_1}=k_{E_2}=k_{E_3}=k_N =1.
\eea
In this regard, the neutrino modular $A_4$ invariant superpotential is given by
\bea
W_{\nu}=g_1~
\Big((N^c H_u L)_{3S}Y_3^{(2)}(\tau_{\nu})\Big)_1+g_2~\Big((N^c H_u L)_{3A}Y_3^{(2)}(\tau_{\nu})\Big)_1+
 \Lambda~ \Big(N^c \otimes N^c)_3 \otimes Y_3^{(2)}(\tau_{\nu})\Big)_1,\label{nusuperpot.}
\eea
where $g_1$ is the coupling constant of the term of the symmetric triplet arising from the product of the two triplets $L$ and  $Y$, while $g_2$ is the coupling of the antisymmetric triplet term. After spontaneous symmetry breaking, the scalar field $H_u$  acquire VEV  $v_u$, the Dirac neutrino mass matrix will be given by the same form in Eq.(\ref{numassmatrix1}), and the right handed neutrino mass matrix is given by
      \bea
 M_R=\Lambda\left(
                    \begin{array}{ccc}
                    2  y_1 & - y_3 & - y_2 \\
    - y_3 & 2 y_2 & - y_1 \\
    - y_2 & - y_1 & 2 y_3 \\
                    \end{array}
                  \right).
\eea

If $\tau_{\nu}=-\frac{1}{2}+i\frac{1}{2}$, the Hermitian Neutrino mass matrix $M_{\nu}$ is invariant under $ST^2ST$ transformation, $$(ST^2ST)^{\dagger}~ M_{\nu} ~ST^2ST =M_{\nu},$$ where
 \bea
ST^2 ST=\frac{1}{3}\left(
                        \begin{array}{ccc}
                         -1 & 2\omega^2 & 2\omega \\
                        2\omega & -1 & 2\omega^2 \\
                       2\omega^2 & 2\omega & -1 \\
                    \end{array}
                 \right).\label{s1 transformation}
\eea


In this case, two eigenvalues of the neutrino mass matrix vanish, so we consider small deviation from the fixed points to get the experimental masses and mixings.



\begin{figure}[htb!]
\centering
\includegraphics[width=7cm,height=7cm,angle=0]{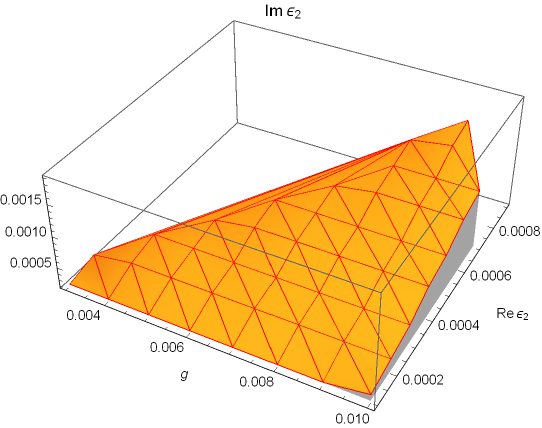}$~~~~~~$

\caption{The allowed values of $g$ and $\delta_2$ for the inverted hierarchy neutrino masses where $\tau_{\ell}=-0.54+ 0.896i$, $\phi=[-\pi/2, \pi/2]$}\label{fig3}
\end{figure}
 Our deviation from fixed points is controlled by small parameters $\epsilon_i$ of order few percent, allowing us to safely ignore higher orders. As previously stated, the exact values of $\epsilon_i$ are determined by the specific potential used to stabilize the moduli. They can also be generated through the running from the high energy scale to the fermion masses scale. Here, we will assume that they are small free parameters. The fixed points above are shifted as $\tau_{\ell}=\tau_1+\epsilon_1$ and $\tau_{\nu}=\tau_{2^\prime}+\epsilon_2$. The parameters are scanned to get the allowed values for which the observed data can be achieved. For $\phi=[-\pi/2, \pi/2]$, $\epsilon_1=0.04+0.03i, ~\frac{\lambda_2}{\lambda_1}=0.0003, ~\frac{\lambda_3}{\lambda_1}=0.06$, the allowed values of $g$ and $\epsilon_2$ are shown in Fig.\ref{fig3}.

In this model, the observed masses and mixing angels are arisen by small deviations from the fixed points $\tau_1,~\tau_{2^\prime}$, so the observed lepton
masses and mixing are consequences of breaking $Z_3$ and $Z_2$ residual symmetries that arisen at $\tau_1$ in charged leptons and $\tau_{2^\prime}$ in neutrino sector respectively.\\

\section{Quark masses and mixing}
The mixing in quark sector is so small relative to the mixing in lepton sector, and the mixing matrix has not special form. The quark mixing matrix can be regarded as a deviation from unit matrix. In this section we assume that the quark mixing is a result of small deviations from the fixed points. We assume the value of quark moduli to be as that in the neutrino sector with small deviations, $$\tau_{q_{\alpha}}= i+\delta_{\alpha}, $$ where $\alpha$ refers to up and down sectors. 
It is crucial to clarify that in our assumptions,
we consider that quarks are associated with a single modulus, denoted as $\tau = i$, with
slight deviations from the fixed point. These deviations can be achieved through radiative
corrections of the stabilizing potential. Therefore, our model continues to rely on three
moduli, as highlighted earlier.

\begin{table}
  \centering
  \begin{tabular}{|c|c|c|c|c|c|c|c|c|c|}
  \hline
  fields & $Q$ &$u_1^c$ & $u_2^c$ & $u_3^c$ & $d_1^c$ & $d_2^c$ & $d_3^c$ \\
   \hline
  $A_4$ & 3 &$1$&$1^{\prime\prime}$&$1^{\prime}$&1&$1^{\prime\prime}$&$1^{\prime}$\\
   \hline
  $k_I$ & 1  & 1& 1&1&1&1&1  \\
  \hline
\end{tabular}
  \caption{Assignment of quarks under $A_4$ and the modular weight $k_I$}\label{quarks}
\end{table}
With the assignments of the quark fermions given in Table \ref{quarks}, the quark invariant
superpotential can be written as
\bea
W_q&=&h^u_1 u_1^c H_u\Big(Q \otimes Y_3^{(2)}(\tau_{up})\Big)_1 +h^u_2 Q_2^c H_u\Big(Q \otimes Y_3^{(2)}(\tau_{up})\Big)_{1^{\prime}}+h^u_3 Q_3^c H_u\Big(Q \otimes Y_3^{(2)}(\tau_{up})\Big)_{1^{\prime\prime}}\nonumber\\
&+& h^d_1 d_1^c H_d\Big(Q \otimes Y_3^{(2)}(\tau_{down})\Big )_1 +h^d_2 d_2^c H_d\Big(Q \otimes Y_3^{(2)}(\tau_{down}) \Big)_{1^{\prime}}+h^d_3 d_3^c H_d\Big(Q \otimes Y_3^{(2)}(\tau_{down})\Big )_{1^{\prime\prime}}.
\label{quarksuperpot.}
\eea
The down and up quark mass matrices are
\bea
m_{\alpha}&=&v_d \left(
                      \begin{array}{ccc}
                        h^{\alpha}_1 & 0 & 0 \\
                        0 & h^{\alpha}_2 & 0 \\
                        0 & 0 & h^{\alpha}_3 \\
                      \end{array}
                    \right)\times \left(
      \begin{array}{ccc}
        y_1 & y_3 & y_2 \\
        y_2 & y_1 & y_3 \\
        y_3 & y_2 & y_1 \\
      \end{array}\right),
      \eea
where $\alpha= u, ~ d$. If $\tau_q=\langle \tau_2 \rangle = i$, the Hermitian matrix $M_{\alpha}=m_{\alpha}^{\dagger} ~m_{\alpha}$ is invariant under $S$ transformation,
$$S^T~M_{\alpha}~ S=M_{\alpha}.$$
For the leading order of approximation, when $\tau_{down}=\tau_{up}=\langle \tau_2 \rangle = i$, the quark mixing matrix $U_{VCM}=V_u^{\dagger}~V_d$ is the diagonal unit matrix. The correct mixing can be achieved by different small deviations of $\tau_{down}$ and $\tau_{up}$ from the value $\langle \tau_2 \rangle = i$. Although both up and down quarks shared the same moduli fixed point, their fluctuations around this fixed point may differ, as these fluctuations depend on the details of each sector.
If \bea
\tau_{down}&=&0.03+1.05 i, ~\tau_{up}=1.06 i,\nonumber\\ ~\frac{h^d_1}{h^d_3}&=&0.027, ~\frac{h^d_2}{h^d_3}=0.0057,~\frac{h^u_1}{h^u_3}=0.00007,~\frac{h^u_2}{h^u_3}=0.0085,
\eea
 the quark mass ratios and the mixing are
\bea
\frac{m_d}{m_b}=0.0011, ~\frac{m_s}{m_b}&=&0.023,~ \frac{m_u}{m_t}=0.000011,~\frac{m_c}{m_t}=0.0076,\nonumber\\
|V_{CKM}|&=& |V_u^{\dagger}~V_d|=
\left(
\begin{array}{ccc}
 0.974 & 0.217 & 0.006 \\
 0.217 & 0.973 & 0.034 \\
 0.0023 & 0.034 & 0.999 \\
\end{array}
\right),
\eea

\section{Conclusion}
In this paper we have investigated $A_4$ modular invariance with three moduli. For each fermion sector, a different modulus has been assigned. We conducted our analysis around the fixed points of these moduli, which allow for residual symmetries. To approximate breaking those residual symmetries and obtain the correct fermion mixing matrices, a small deviation from the moduli fixed points is required in some cases. We presented specific benchmark points that provide the best fit of the masses and mixing of quarks, charged leptons, and neutrinos.

 \end{document}